\documentclass[aas2pp4,natbib209]{emulateapj}

\usepackage{epstopdf}
\usepackage{natbib}
\usepackage{graphicx}

\slugcomment{2017 Mar 9}
\newcommand{\msun}{{\rm M_\odot}}
\newcommand{\tco}{\ifmmode {^{13}{\rm CO}} \else {$^{13}{\rm CO}$}\fi}
\newcommand{\dco}{\ifmmode {^{12}{\rm CO}} \else {$^{12}{\rm CO}$}\fi}
\newcommand{\juz}{\ifmmode {{\rm J}=1\rightarrow 0} \else
{J=1$\rightarrow$0}\fi}
\newcommand{\jdu}{\ifmmode {{\rm J}=2\rightarrow 1} \else
{J=2$\rightarrow$1}\fi}
\newcommand{\jtd}{\ifmmode {{\rm J}=3\rightarrow 2} \else
{J=3$\rightarrow$2} \fi}

\begin{document}


\title{Planet Formation in AB Aurigae: Imaging of the inner gaseous Spirals observed inside the Dust Cavity}
\author{Ya-Wen Tang$^1$, Stephane Guilloteau$^2$, Anne Dutrey$^2$, Takayuki Muto$^3$, Bo-Ting Shen$^4$, Pin-Gao Gu$^1$, Shu-ichiro Inutsuka$^5$, Munetake Momose$^{6}$, Vincent Pietu$^7$, Misato Fukagawa$^8$, Edwige Chapillon$^7$, Paul T. P. Ho$^1$, Emmanuel di Folco$^{2}$, Stuartt Corder$^9$, Nagayoshi Ohashi$^{10}$, and Jun Hashimoto$^{11}$}
\affil{$^{1}$Academia Sinica, Institute of Astronomy and Astrophysics, Taipei, Taiwan}
\affil{$^{2}$Laboratoire d'astrophysique de Bordeaux, Univ. Bordeaux, CNRS, B18N, allŽe Geoffroy Saint-Hilaire, 33615 Pessac, France}
\affil{$^{3}$Department of Physics, National Taiwan University, Taiwan}
\affil{$^{4}$Division of Liberal Arts, Kogakuin University, 1-24-2 Nishi-Shinjuku, Shinjuku-ku, Tokyo 163-8677, Japan}
\affil{$^{5}$Department of Physics, Graduate School of Science, Nagoya University, Furo-cho, Chikusa-ku, Nagoya 464-8602, Japan}
\affil{$^{6}$College of Science, Ibaraki University, 2-1-1 Bunkyo, Mito, Ibaraki 310-8512, Japan}
\affil{$^{7}$IRAM, 300 rue de la Piscine, Domaine Universitaire, 38406 Saint-Martin-d'H\`{e}res, France}
\affil{$^{8}$Division of Particle and Astrophysical Science, Graduate School of Science, Nagoya University, Furo-cho, Chikusa-ku, Nagoya, Aichi 464-8602, Japan}
\affil{$^{9}$National Radio Astronomy Observatory, 520 Edgemont Road, Charlottesville, VA, 22903, USA}
\affil{$^{10}$Subaru Telescope, National Astronomical Observatory of Japan, 650 North AÕohoku Place, Hilo, HI 96720, USA}
\affil{$^{11}$Astrobiology Center of NINS 2-21-1, Osawa, Mitaka, Tokyo, 181-8588, Japan}

\email{ywtang@asiaa.sinica.edu.tw}

\begin{abstract}
We report the results of ALMA observations of a protoplanetary disk surrounding the Herbig Ae star AB Aurigae.  We obtained high-resolution (0$\farcs$1; 14 au) images in $^{12}$CO (J=2-1) emission and in dust continuum at the wavelength of 1.3 mm. 
The continuum emission is detected at the center and at the ring with a radius of $\sim$ 120 au.  
The CO emission is dominated by two prominent spirals within the dust ring. 
These spirals are trailing and appear to be about 4 times brighter than their surrounding medium. 
Their kinematics is consistent with Keplerian rotation at an inclination of $23\degr$. 
The apparent two-arm-spiral pattern is best explained by tidal disturbances
created by an unseen companion located  at 60--80 au, with dust confined in the pressure
bumps created outside this companion orbit. 
An additional companion at r of 30 au, coinciding with the peak CO brightness and a large pitch angle of the spiral, would help to explain the overall emptiness of the cavity.
Alternative mechanisms to excite the spirals are discussed. The origin of the large pitch angle detected here remain puzzling.

\end{abstract}

\section{Introduction}

Recent large facilities such as ALMA, HiCIAO/SUBARU or SPHERE/VLT have revealed complex gas and dust structures, such as large cavities, asymmetries and spiral patterns in protoplanetary disks, thanks to their high sensitivity and resolving power. 
The spiral-like features, either observed in the millimeter (mm) and sub-mm wavelengths or in the optical and infrared (IR) scattered light, are found around, for example, HD163296
\citep{Grady+etal_2000,Fukagawa+etal_2010},  HD100546 \citep{Grady+etal_2001,Quanz+etal_2011},
HD142527 \citep{Fukagawa+etal_2006,Christiaens+etal_2014}, HD97048 \citep{Doering+etal_2007},
MWC758 \citep{Isella+etal_2010,Grady+etal_2013,Benisty+etal_2015},
HD135344B \citep{Muto+etal_2012}, Elias 2-27 \citep{Perez+etal_2016}, and 
HD\,141569A \citep{Clampin+etal_2003,Perrot+etal_2016}.

These spirals may appear reasonably regular, asymetric or more reminiscent of faint
discontinuous disks. The origin is basically due to gravitational disturbances, either
via planet-disk interactions \citep[e.g.][]{Zhu+etal_2015} or gravitational
instabilities inside a massive disk \citep{Kratter+etal_2016}.  
The resulting structures can be very complex \citep{Dipierro+etal_2014,dong2015,Flock+etal_2015,Pohl+etal_2015}.
Nevertheless, spiral-like features are expected to reveal disturbers, such as
embedded planets,  which cannot be yet directly observed when the gas of the disk is not fully dissipated.

Located at 140~pc, AB Aurigae (hereafter AB Aur) is one of the closest Herbig Ae stars of spectral type of A0
Ve \citep[2.4$\pm0.2\msun$,][]{dewarf2003}  and it shows an extended
reflexion nebula at large scale (100s au). This young star is unique, because it exhibits a very small inner
disk ($\sim 2-5$ ~ au) observed in near-IR (NIR) and mid-IR interferometry  \citep{DiFolco+etal_2009},
surrounded by a large cavity of radius $\sim 70$\,au based on the CO
and mm continuum observations \citep{Pietu+etal_2005}, and a CO gas and dust rotating ring from which external spirals
are observed both in the NIR \citep{Fukagawa+etal_2004} and in the mm domain
\citep{tang2012}.  The inclination of the inner and outer disks slightly varies from 23$\degr$ at 20
au to 29$\degr$ at 100 au scale \citep{tang2012}. The accretion rate observed for AB
Aur \citep[1.3$\times~10^{-7}~\msun~yr^{-1}$;][]{salyk2013,garcia2006} is still high given
the estimated age of the star (about 4 Myr).

The large scale spirals detected at optical/NIR wavelengths
\citep{Grady+etal_1999,Fukagawa+etal_2004} extend from 100 au up to 500 au.  The
extended CO emission mapped by the IRAM array also reveals large scale spirals.
\citet{tang2012} found that surprisingly, the excess of CO gas along the spirals is
apparently counter-rotating with respect to the gaseous disk, that \citet{tang2012}
attribute to residual gas being accreted from the envelope high above the disk mid-plane.
Such a scenario is also reminiscent of the asymmetric accretion suggested for some Class
0 sources by \citet{Tobin+etal_2011}.

At small scales, the wide dust gap, the warped disk (change in inclination from small to
large scales by about  6-10 degrees) and the asymmetric dust ring (an intensity contrast
of 3 is observed inside the mm dust ring), suggest the existence of, at least, one
undetected companion with a mass of 0.03 $\msun$ at a radius of 45 au \citep{tang2012}.

Motivated by such complex structures, we obtained ALMA observing time in Cycle 3 to image
at very high angular resolution ($< 0.1''$) the close environment of  AB Aur in CO and in
mm continuum. In this paper, we report the observations (Section 2) and the results (Section
3) of these data and then discuss them with respect to recent studies and models of
planet-disk interactions (Section 4). For simpler comparison with previous work,
we assume a distance of 140\,pc, although GAIA DR1 release indicates $153\pm10$ pc
\citep{GaiaDR1_2016}.

\section{Observations}
The observations were carried out with ALMA in Cycle 3 (project "2015.1.00889.S") with
two execution blocks on Nov. 5 and Nov. 10, 2015, respectively. The array included 45 and
46 of the 12-m antennas, respectively. The baselines ranged from 78.1 m to
15.5 km. The gain calibrator (phase) is J0512+2927, which is 3.83 degree away from AB
Aur and was observed with a cycle time between AB Aur and J0512+2927 of 1 minute. The
bandpass calibrator was J0510+1800. The absolute flux level is referenced to J0423-0120
and J0510+1800 for which flux densities of  0.77 Jy and 3.42 Jy, respectively, were determined by
the ALMA monitoring program. The flux calibration is expected to be accurate at the 10\% level.

The calibrations were done by EA ARC using CASA 4.5. A secondary calibrator,
J0518+3306, was observed every 15 minutes in order to verify the
position accuracy. It lies 5.38 degree Eastwards of AB Aur, and is
3.83 degree away from the phase calibrator J0512+2927.  After standard calibration, the
position difference between the known coordinate and the apparent position of
the secondary calibrator is 0$\farcs$019 and 0$\farcs$036 for the two execution blocks.
We conclude that the absolute positioning accuracy of this experiment is limited
to $\sim 0\farcs$03.

After calibration, the data were exported to the GILDAS package for further
analysis. 
The detected central continuum peak of AB Aur at 1.3 mm
is offset by (-5$\pm0.3$, -10$\pm0.4$) mas from to the stellar location, 
taken from the Hipparcos reduction of \citet{2007Leeuwen}.
As this offset is of the same order as seen for the secondary calibrator, it is
likely due to the absolute position uncertainties. 
We assume the continuum peak and the stellar positions are identical.

The presented data have been corrected for the proper motion with value of (2.63,
-24.73) mas yr$^{-1}$ given by \citet{2007Leeuwen}, and shifted to the epoch of 2000.0 for
comparison with earlier work. All the presented maps are centered on the 1.3\,mm continuum
peak, which is at ($\alpha$, $\delta$) = (04:55:45.85, 30:33:04.30). The maps of
$^{12}$CO 2-1 are with spectral resolution of 0.71 km\,s$^{-1}$ and sensitivity of 
1.60 mJy per 0$\farcs$11$\times$0$\farcs$08 beam (4.6 K). 
The effective sensitivity
(1$\sigma$, including dynamic range limitations) of the continuum emission is 33 $\mu$Jy per 0$\farcs$14 beam (0.04 K).

%
%
\begin{figure*}
\begin{center}
\includegraphics[scale=1.1]{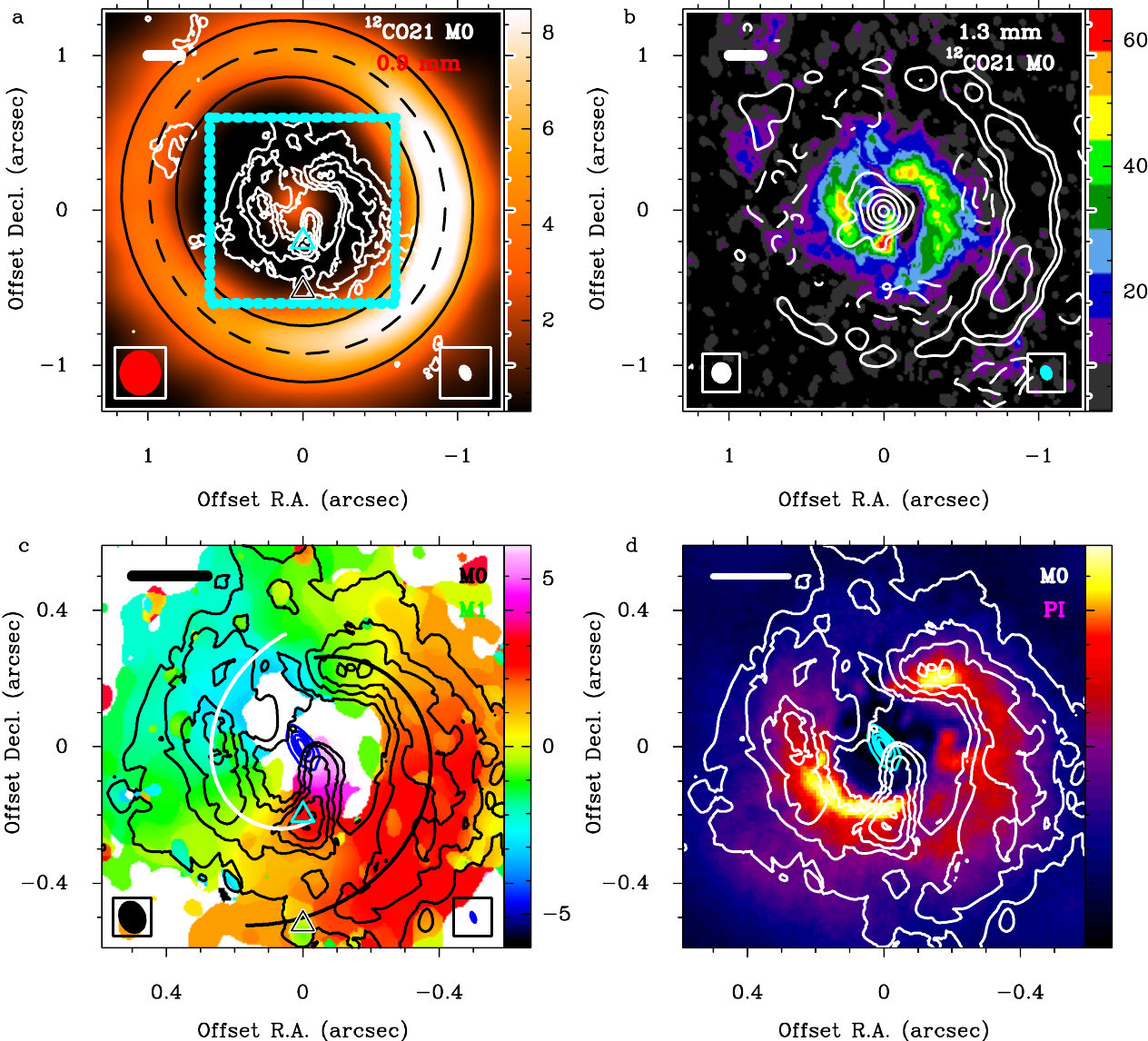}
\caption{
(a): Moment 0 map of CO 2-1 (contours) and 0.9 mm continuum (color scale; unit in mJy per 0$\farcs$3 beam; Tang et al., in prep).
The cyan box marks the zoom-in area shown in panel c and d.
The triangle marks the possible locations of the planets. 
(b): Continuum emission at 230 GHz (1.3 mm) (contours) and the moment 0 map of 
CO (color scale; unit in mJy/beam km/s, where the beam is 0$\farcs$11).
The contours are -6, -3, 3, 15, 30 and 50$\sigma$, where 1$\sigma$ is 33 $\mu$Jy per 0$\farcs$14 beam (0.04 K).
(c) Moment 0 (black contours) and moment 1 with $v_{\rm sys}$ subtracted (color scale; unit in km/s) maps of CO with an angular resolution of 0$\farcs$11. 
Continuum emission at 1.3 mm with 0$\farcs$03$\times$0$\farcs$07 beam (marked in the lower right corner) is shown in thin blue contours at 10, 20, 30 and 40$\sigma$, where 1$\sigma$ is 27 $\mu$Jy/beam. The white and black arcs mark the best-fit spirals.
(d) Moment 0 map of CO (white contours) and polarized intensity image at NIR
\citep[color scale, arbitrary unit; adopted from][]{Hashimoto+etal_2011}. 
The cyan contours are the same as in panel c.
A scale bar of 30 au is marked at each upper-left corner.  
In panel a,c and d, the black contours are 2,4,6,8 and 10$\times$ 6.15 (mJy/beam km/s). 
The angular resolution of the color image and contour image is marked as an ellipse at the lower-left and lower-right corner, respectively, in panel a,b, and c.
}\label{Fig:6panels}
\end{center}
\end{figure*}

\section{Results}

\subsection{Continuum Emission}
The continuum emission at 1.3 mm is detected at 60$\sigma$ level toward the center, and
clearly detected at the dust ring (Fig. \ref{Fig:6panels}b). The total detected flux
density is 16.4 mJy, which is 20\% of the reported flux density of 80 mJy at 1.3 mm from
our previous detection with shorter baselines \citep{tang2012}. This suggests that
the missing 80\% of the flux is at the scale larger than 2$\farcs$1, which is determined
by the shortest baseline of our observation. 
This missing flux at a scale larger than 2$\farcs$1 corresponds to a maximum brightness of the dust ring of $\sim 0.2$ mJy per $0\farcs14$ beam. 
Accounting for this missing flux, the contrast in the dust ring between the brightest regions
in the South-Eastern part and the rest can be as low as about 3, although
Fig. \ref{Fig:6panels}a displays an apparent contrast around 5. 
The dust ring is consistent
with a circular structure of radius $\sim 120$ au, inclined at $\sim 23\degr$ with the
same axis as derived from the gas kinematics by \citet{Pietu+etal_2005}, and centered
on the star. More precise values of the ring parameters are actually derived
from higher sensitivity, lower angular resolution observations at 0.9 mm (Tang et al., in prep).



We model the central continuum peak with an elliptical disk in the
visibility plane. 
The best fit disk diameter is found to be $\sim 11$\,au.
The flux density is 2.1 mJy, which is larger than the previous detected value of 1.3
mJy with the IRAM interferometer \citep{tang2012}. 
The difference is at $3 \sigma$ level when accounting for the flux calibration accuracy. 
Furthermore, the different spatial filtering properties of the two observations may also contribute to the measured differences, especially since the IRAM observations did not clearly separate the central continuum from the dust ring emission.
We note that the total central continuum emission is about 3.4 mJy within the inner 0$\farcs$6 region and the emission appears extended at the level of SN ratio of 3. 

The nature of the central continuum emission is unclear. 
There is an elongated emission at 3.3 cm detected with SN ratio of 4 along PA of $\sim$170$\degr$ and was suggested as free-free jet by \citet{rodriguez2014}. 
The total flux density of the jet is reported to be 0.8 mJy at 7 mm.
Assuming a spectral index of 0.6 for the jet, this free-free emission would give 2.2 mJy at 1.3 mm within the central 1$\arcsec$. 
About 65\% of the emission at 1.3 mm (2.1 mJy from the inner disk and
1.3 mJy from the faint emission) can come from free-free. 


%
%
%
%
\begin{figure*}[!ht]
\includegraphics[scale=0.6]{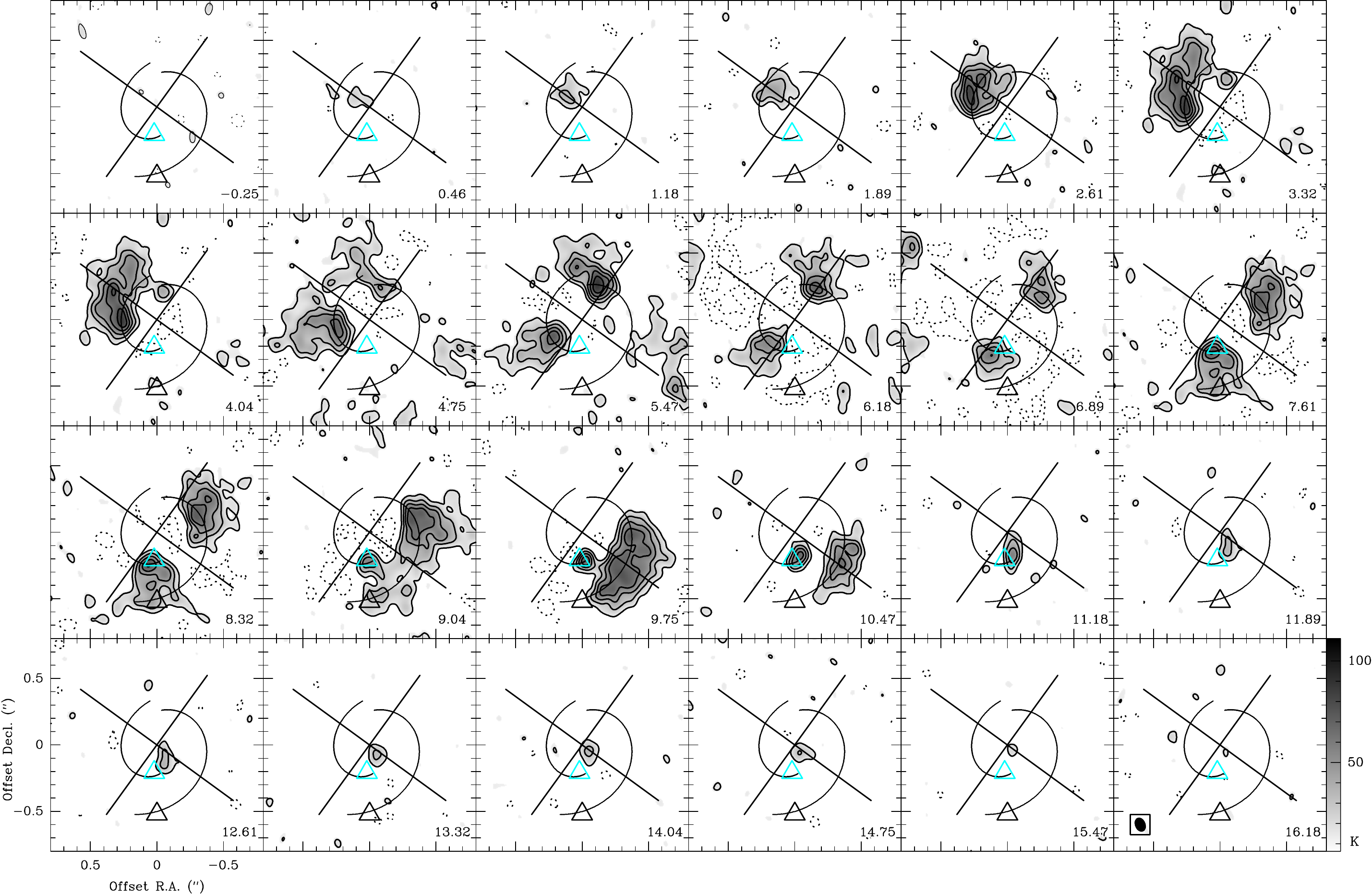}
\caption{Channel maps of $^{12}$CO 2-1. The contours denote the brightness temperature
from -6,-3,3 to 15 by 3$\sigma$, where 1$\sigma$ is 4.6 K (or 1.6 mJy per 0$\farcs$11$\times$0$\farcs$08 beam).
The segments mark the major (P.A. of 54$\degr$) and minor axis (P.A. of -36$\degr$) of the disk.
$v_{\rm LSR}$ is labelled at the lower-right corner of each panel.}
\label{Fig:12co-p1-chan}\label{Fig:12cochan}
\end{figure*}

\subsection{CO Gas}
\label{sub:co}

Previous observations of AB Aur in CO 2-1 and 3-2 at angular resolutions
of $\sim 0$\farcs$5-1\arcsec $ \citep{Pietu+etal_2005,Lin+etal_2006} revealed that most
of the gas was in a large outer disk surrounded by an extended envelope \citep{tang2012}.
The disk displayed an inner radius of 45 $-$ 77 au \citep{Pietu+etal_2005} depending
on the isotopologue being observed, although some $^{12}$CO emission was associated with
the unresolved inner dust disk \citep{tang2012}.
Large scale CO ``spirals'', extending outside the dust ring into the envelope, were reported
in \citet{Lin+etal_2006} and \citet{tang2012}. Their kinematics suggest that these
features trace gas infalling onto the ring from well above and below the disk mid-plane
\citep{tang2012}. On the other hand, a recent analysis of the NIR images, which uses
the spectral-energy-distribution fitting, argue that these large scale spirals can be disk-related \citep{2016Lomax}.
In any case, all structures reported in these earlier studies are at r$>$100 au and beyond.

In contrast, the new high angular resolution (about 0$\farcs$1) and high
sensitivity (noise of 2.26 K) $^{12}$CO 2-1 images obtained with ALMA (Fig.
\ref{Fig:6panels}) reveal that most of the detected emission here is  {\it inside} the
dust cavity, within 90 au from the star. The $^{12}$CO 2-1 emission at various velocity $v_{\rm LSR}$ (i.e. channel maps) is shown in Fig. \ref{Fig:12cochan}. Fig.\ref{Fig:6panels} clearly shows that the CO gas is inside the
dust ring and there is no apparent connection with the large CO disk and the extended CO
spirals mentioned in the previous works. This apparent lack of connection may, however, be
partly an artefact due to our $uv$ coverage. Any extended and smooth emission larger than
about 2$\farcs$1 would have been filtered out. Only compact CO features, such as spirals
or rings, remain detectable.

Some idea of the impact of this filtering can be obtained by comparing 
with the results from Tang et al. (2012), which included short spacings from a 30-m 
single dish map. 
Smoothing our data to the same angular resolution 
as in Tang et al 2012 (0$\farcs$56 $\times$ 0$\farcs$42) shows that we are only recovering 
25 \% of the line flux in the velocity ($v_{\rm LSR}$) range of 2 to 5 and 7 to 10 km/s. 
Near the systemic velocity ($v_{\rm sys}$ of 5.85 km/s) in $v_{\rm LSR}$ of 5 to 7 km/s, the missing flux 
problem is more severe.  
Given the typical brightness measured by \citet{tang2012} of 50 K, we estimate that the measured 
brightness (at the velocity where the emission peaks) should be 
de-biased by adding about 20 to 30 K, as a first order correction
for missing flux.

%
%
\begin{figure*}[!th]
\includegraphics[scale=0.7,trim={1.5cm 0 2.0cm 0},clip]{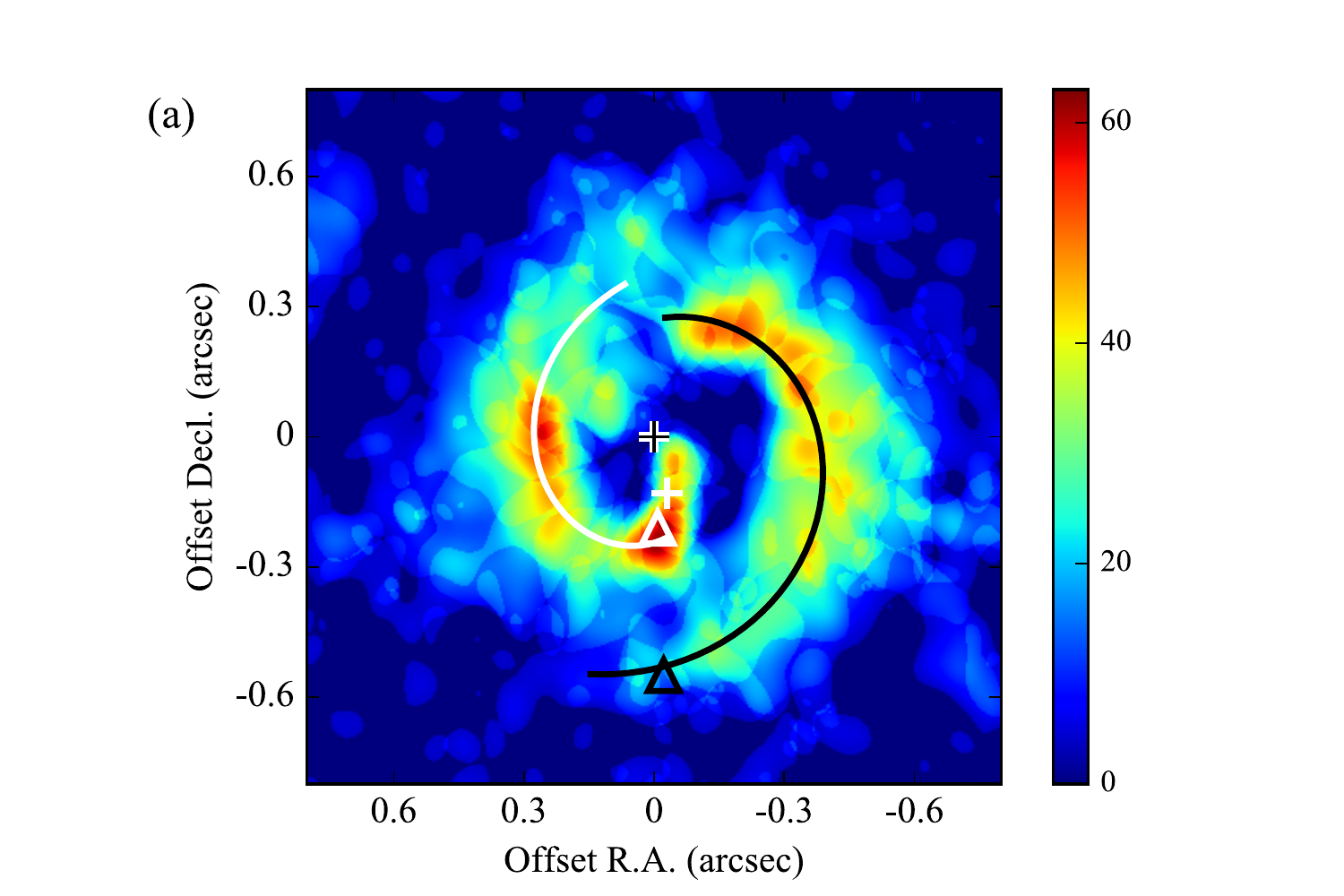}
\includegraphics[scale=0.7,trim={0.5cm 0 0 0},clip]{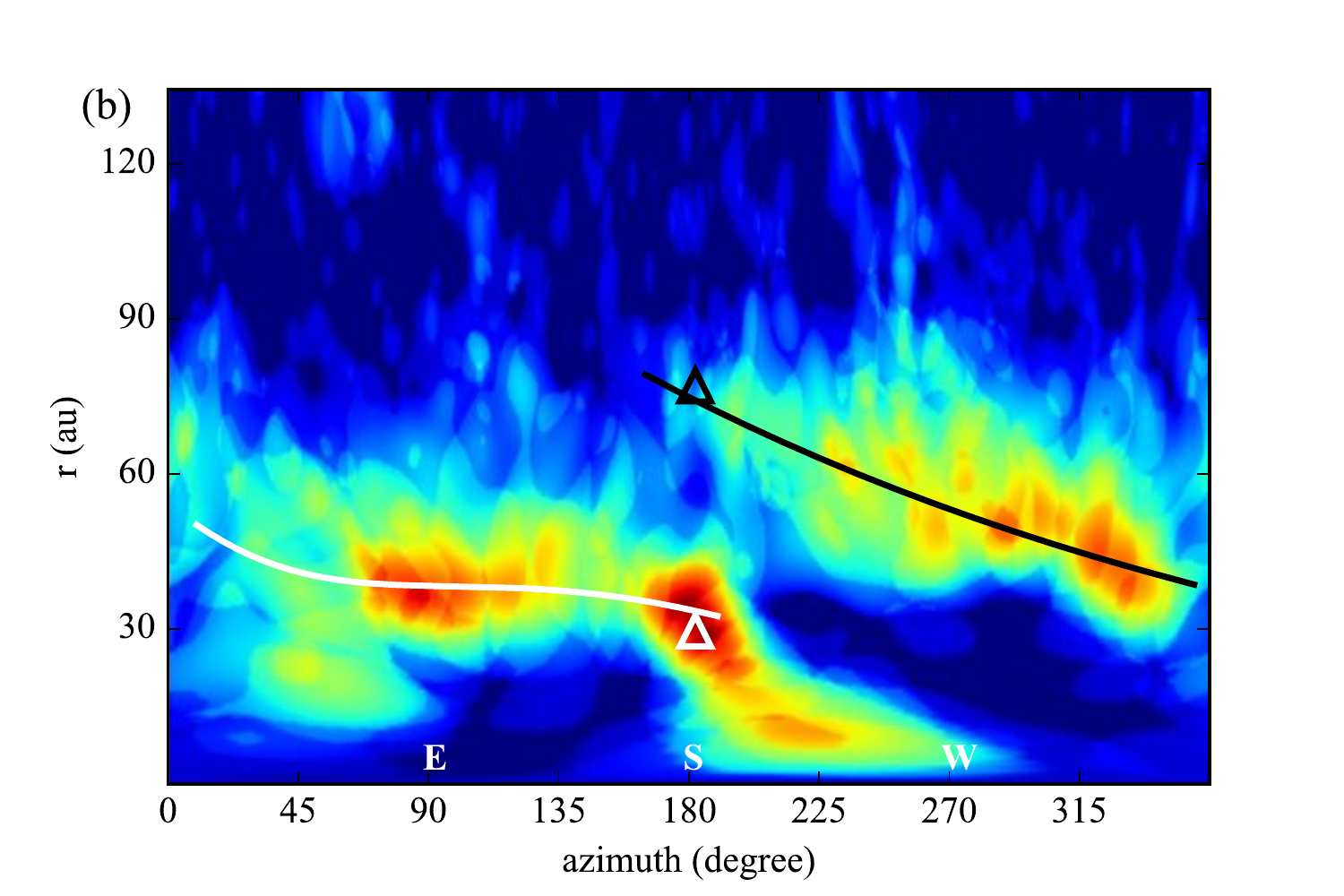}\\
\includegraphics[scale=0.55,trim={3cm 13cm 2cm 1cm},clip]{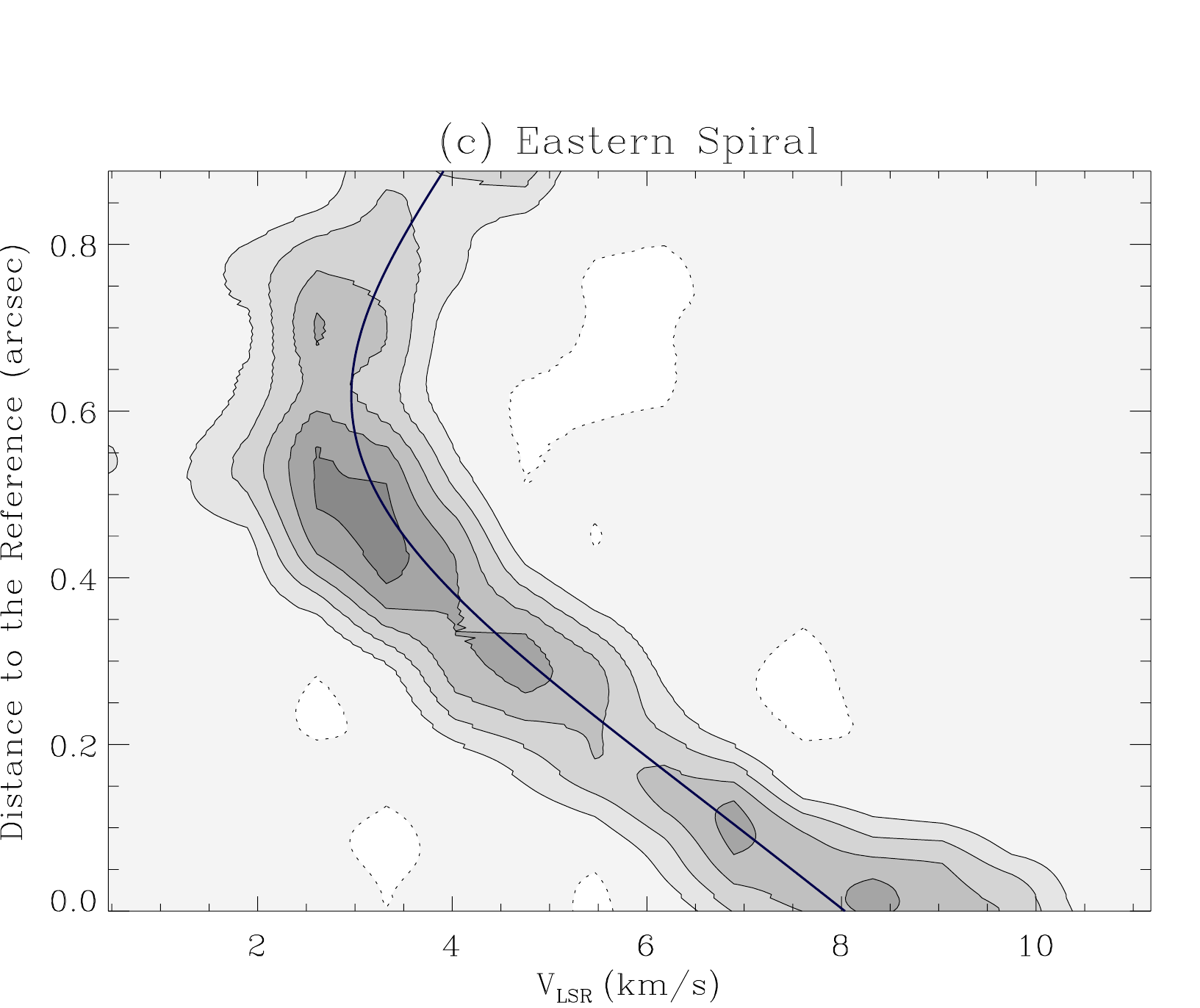}
\includegraphics[scale=0.55,trim={3cm 13cm 2cm 1cm},clip]{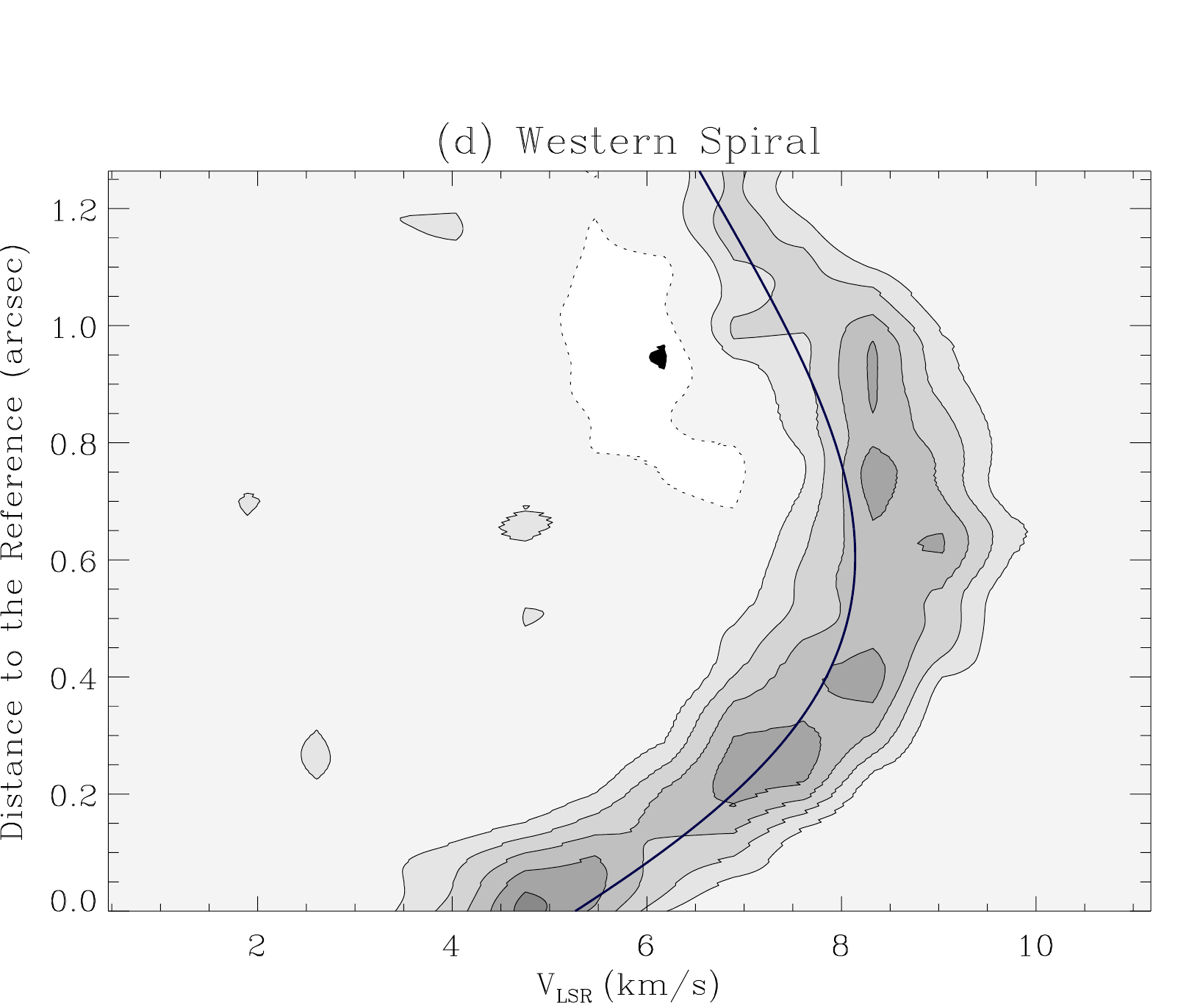}
\caption{(a): De-projected moment 0 map of $^{12}$CO 2-1 emission (in mJy/beam km/s). The arcs and pluses mark the best fit spiral and the origins of the best-fit in corresponding colors. 
The triangles mark the possible locations of the
planets.  
(b): De-projected moment 0 map of $^{12}$CO in polar coordinate. The rest of the symbols are the same as in panel a.
(c) and (d): Position-velocity plots of $^{12}$CO 2-1 along the eastern and
western spirals marked in panel a. The x-axis is $v_{\rm LSR}$ in units of km/s, and the
y-axis is the distance to the starting point of the arm, which is the near end toward the
star of the arm, in units of arcsecond. The contours are -3, 3, 6, 9, 12 and 15$\sigma$, where 1 $\sigma$ is 0.16 mJy per 0$\farcs$11 beam. The curves mark the expected Keplerian velocity
with stellar mass of 2.4 $\msun$ and an inclination angle of 23.2$\degr$ (solid line)
}
\label{Fig:comparison}
\end{figure*}

\subsubsection{General Features}

The detected $^{12}$CO 2-1 gas emission peaks at position P1 (see Fig. \ref{Fig:6panels}),
about 30 au from the star, where the peak brightness temperature, $T_{\rm B}$, of 70 K at $V_{\rm LSR}$ of 8.32 km/s. 
At this distance, following
\citet{Pietu+etal_2005}, we assume CO traces gas at a temperature of 140 K.  With a
linewidth of 1 km\,s$^{-1}$ and optically thin emission, the column density of CO gas is
10$^{17}$ cm$^{-2}$ (surface density, $\Sigma_{\rm gas}$, 10$^{-2}$ g/cm$^2$) for $T_{\rm B}$
of 70 K, which,  for a CO gas abundance of 10$^{-4}$ with respect to H$_2$, gives
a total gas mass of 0.03 $M_{\Earth}$ in one beam. As the CO gas emission is at least
partially optically thick, this gas mass is only a lower limit. The non-detection
of C$^{18}$O 3-2 (3$\sigma$ upper limit) indicates an
upper limit for the mass 180 times larger (abundance ratio of CO to C$^{18}$O of 560), being 5.4 $M_{\Earth}$.

The integrated intensity (moment 0) map of the $^{12}$CO 2-1 emission appears asymmetric
with respect to the central star (Fig. \ref{Fig:6panels}). The emission is patchy,
especially in the fainter part, but appears to the first order spiral-like. We separate
the CO emission into an eastern arm and a western arm for further analysis. The total gas
mass of the eastern and western spirals is estimated to be in the order of 1 $M_{\Earth}$
or more.

To display the radius dependence of these structures, we present in Figure
\ref{Fig:comparison}b the deprojected moment 0 map (see description below) in polar coordinates ($r,\theta$), where $r$ is
the distance from the 1.3 mm continuum peak, and $\theta$ is the Azimuth counted eastwards from North.
Both spirals appear to have a smaller $r$ as azimuth increases, and are thus trailing
given the known sense of rotation of the AB Aur disk. The eastern spiral
consists of two separated structures, with the structures around azimuth of 180$\degr$
having a clearer $r$ dependence.

We further analyze the CO emission as follows. Firstly, the moment 0 map is deprojected
with an inclination of 23.2$\degr$ assuming the P.A. of the disk rotation axis of -36$\degr$ and
centered on the 1.3 mm continuum peak (Fig. \ref{Fig:comparison}a). 
We then take the maximum locations of the $^{12}$CO moment 0 image every 5$\degr$ in azimuth angle, $\theta$.
{We first fit the eastern and western arm with logarithmic spirals in order to search for the best description.}
For the eastern spiral, the best-fit function is $r(\theta)=0\farcs85 \exp(-21.0\theta)$, where $\theta$ is in degree, and the origin is at (-0$\farcs$01,-0$\farcs$09) from the continuum peak. 
Note that the data at r$<$ 30 au in the eastern spiral is clearly deviating from the best-fit function. In all the figures the best-fit function of the eastern spiral is not shown at r$<$30 au.
For the western spiral, the best fit function is $r(\theta) = 0\farcs38\exp(-12.5 \theta)$ and origin is
at the central continuum peak. The results are marked in Fig. \ref{Fig:comparison}.

\subsubsection{Kinematics}
The highest velocities at which CO gas is detected are $v_{\rm LSR}$ of 0.46 and
15.47 km/s for the blue-shifted and red-shifted gas (see Fig. \ref{Fig:6panels}a),
respectively. We note that the position centroid of the emission at these high
velocities is consistent (within the uncertainty) with the location of the central
continuum peak detected at 1.3\,mm. Because the peak locations of the high velocity CO
gas appear along the disk plane, this suggests that the high velocity gas traces the
inner CO disk in rotation instead of the jet observed at cm wavelengths by
\citet{rodriguez2014}.

We extracted the position-velocity plots along the best-fit spirals (Figs.
\ref{Fig:comparison}c and \ref{Fig:comparison}d). The predicted velocity due to Keplerian
rotation ($v_{\rm kep}$) is indicated in solid line for the nominal inclination of
23.2\degr derived from the dust, a stellar mass, $M_{\star}$, of 2.4 $M_{\sun}$ and P.A. of the
rotation axis of -36$\degr$, using the systemic velocity, $v_{\rm sys}$, of 5.85 km\,s$^{-1}$ from
\citet{tang2012}. 
There is no radial motion detected within the resolution of the reported observations (0$\farcs$1 angular resolution and 0.7 km/s spectral resolution).

\subsection{Comparison of ALMA observations with the optical/NIR images}

Within the mm dust cavity, small dust grains ($\mu$m size) have been detected in
scattered light in the optical and NIR \citep{Grady+etal_1999,
Fukagawa+etal_2004}. The polarized intensity (PI) image at 1.6 $\mu$m by
\citet{Hashimoto+etal_2011} exhibits asymmetric features as a function of azimuth. The
observed scattered light emission traces warm small dust particles at the top of the disk
surface. \citet{Hashimoto+etal_2011} fitted two inclined rings and one ring gap to the PI
emission. However, the derived orientations of the rings and the gap deviate from each
other, suggesting that the actual structures are not rings. The CO gas distribution from
our new results shows structures which are similar to those seen in the PI image (see
Fig. \ref{Fig:6panels}d). Both trace the same spirals, but the 1.6 $\mu$m features are
inwards of the CO structures. We note that the small $\mu$m dust particles responsible
for the 1.6 $\mu$m scattering should be well mixed with the gas, as observed in the case
of IRS 48 by \citet{2013Marel}. In addition, the extended CO emission is filtered out by
ALMA.

\section{Discussions}


Several mechanisms can lead to spiral-like patterns. However, the AB Aur disk density is
not large enough to make it gravitationally unstable at all scales
\citep{Pietu+etal_2005}. The observation of spirals within the dust cavity rather points
towards tidal disturbances by a compact object. We investigate here to what extent an
embedded (planetary mass) object can explain the observed structures. Such an object
located inside the dust cavity may create both the observed mm cavity and spiral-like
pattern \citep[for example,][]{Zhu+etal_2015}.

In \citet{dong2015}, the adiabatic model with a 6 $M_{\rm J}$ planet orbiting at
50 au of $M_{\star}$= 1 $M_{\sun}$ seems to produce structures similar to what we
observe here in $^{12}$CO. Scaling to the mass of AB aur ($M_{\star}$ of 2.4 $M_{\sun}$),
the planet would need to have a mass of 14 M$_{\rm J}$ and be located at $r$ of 30 au (location P1, see section 3.2.1; figure \ref{Fig:6panels}c), in order to create the observed features if we adopt this model.

This possibility is supported by several unusual features detected near the
location of this putative inner planet at P1. First, the pitch angle of the eastern
spiral becomes large near this point: large pitch angles are only obtained near the
driving object in planetary-induced spirals. Second, there is a compact CO emission
between at velocities $v_{\rm LSR}$ of 9.04 to 11.18 km/s (see Fig.
\ref{Fig:12cochan}) and the linewidth is larger than elsewhere, as expected for the streaming
motions (gas accretion and jet at the circumplanetary disk) around the planet  \citep[see][, for example]{Gressel+etal_2013}. Finally, there is 1.3\,mm continuum emission with flux of 110
$\mu$Jy/beam at the location P1. This matches the predicted flux densities of the
circumplanetary disks of known planet candidates, of the order of 100 $\mu$Jy at 1.3\,mm
\citep{2016Zhu}. However, because the 3.3\,cm continuum flux is 35.9 $\mu$Jy/beam at this
position, we can not rule out that the 1.3 mm continuum emission has a contribution from
free-free emission.

However, this scenario faces some difficulties. In the analysis of the multi-epoch NIR images,
\citet{2016Lomax} found no apparent rotation of the spiral patterns within the 5.8 year
time span of the images. If the spiral patterns are induced by a planet or companion,
this suggests that such an object is located  at $r >$ 47 au. However, their first epoch
data, using Stokes I only, is not sensitive to the region around P1 discussed here, so
this result only applies to the other arms. Furthermore, a planet at P1 cannot simply
explain the western spiral, because outwards of the orbit of the
driving planet, the perturbation is mostly a (usually tightly wound) one arm spiral.

On the contrary, two-arm spirals are easily produced inside the orbit of the perturbing
planet \citep[see][, for example]{Fung2015}. In this second scenario, another
possible location of the planet, P2, is at the outer-tip of the western spiral
(i.e. at P.A. of 180$\degr$ and $r = 80$ au). In this case, the spirals seen here are 
inner spirals, and the two arms are explained by a single perturber.
A planet at this location can also explain the sharp inner edge
of the dust ring seen at 1.3\,mm, while there is still a significant amount of
$\mu$m size dust particles \citep{Hashimoto+etal_2011} within the mm dust cavity. This is
consistent with the picture of dust filtration, where dust grains larger than 0.1 mm are
trapped at the radius outside of the planet and only small dust grains and gas can
accrete toward the inner disk \citep{2012Zhu}. This location is also far enough to be
consistent with the lack of apparent rotation of the NIR spiral patterns found by \citet{2016Lomax}.

The main difficulty in this scenario is its failure to explain the large pitch angle between P1 and the star location. 
One possibility would be that the inner disk part of the system is warped compared to the
outer part. Indeed, attributing a single inclination to the AB Aur system appears
challenging. The inclination angle of the dust ring is 23$\degr$, while NIR
interferometric results suggest an inclination of only 20$\degr$ for the inner dust
disk \citep{Eisner+etal_2004}. 
The larger scale observations from \citet{Pietu+etal_2005} suggest higher inclinations,
but fail to reproduce the expected Keplerian velocities.
It appears that the variation of inclination as a function of scale can not explain the large pitch angle.

All the spirals induced by planets in simulations appear to have limited contrast, the
maximum being 2:1 in \citet[][]{dong2015}, for example.  
The $^{12}$CO spiral is detected at about 12$\sigma$ level (see figure 3c and 3d). 
Using a 3$\sigma$ non-detection of the surroundings, the observed spirals have a contrast of 4:1, much higher than predictions. 
However, the missing flux (see section 3.2) can easily bring down this contrast to 2:1.
Furthermore, the models predict density contrasts, while we observe a brightness contrast in the CO J=2-1 line.

The simplest interpretation of the observed contrast is linked to the vertical thermal
gradient in the disk and the expected CO opacity.
The spirals have larger column densities, and the $\tau=1$ layer is reached at higher heights in the spirals. 
Therefore, the spirals are warmer gas and far from being dense enough to affect the local scale height. 
We note that from a comparison between $^{12}$CO and $^{13}$CO,
\citet{Pietu+etal_2005} indicated a temperature difference of about a factor of 2 between
the lower layer traced by $^{13}$CO and the $^{12}$CO surface. 
Beside, the near-IR opacity towards the star is much larger than the CO line opacity along our line of
sight. We expect the scattered light to trace the inner side of the spiral structures,
exactly as seen in Fig. \ref{Fig:6panels}d.

These spiral-like patterns are reminiscent of the patterns expected for 
planet-induced structures in disks, although the pitch angle seems 
relatively large (about 20$^\circ$ for the Western spiral). Indeed, it 
is difficult to fit the analytical shape prescription of \citet{Rafikov2002,Muto+etal_2012} 
to our measurement. 
For the south part of the Eastern spiral, we find a planet location 
near 28 au at PA 180$^\circ$ (i.e. near the brightest spot) but the 
best fit values for parameters $h/r$ at the planet radius and $\beta$ 
(the exponent of the velocity law of the sound speed) are 0.4 and 0.9, respectively. This is 
inconsistent with the measured radial temperature profile and expected 
thickness of our disk, as derived from hydrostatic equilibrium. 
The Western spiral analysis lead to similar difficulties, although a
solution with a planet near 20 au is possible assuming $\beta =0$.

Another exciting mechanism of the spiral wave is suggested by \citet{Montesinos+etal_2016}, where
shadows cast by a tilted inner disk onto the outer disk will cause a drop in illumination and thus lower temperature locally. 
The pressure imbalance will further cause a density enhancement.
As suggested, this mechanism could create spiral waves with an opening angle 10-15$\degr$ in the outer disk. 
We note that the CO spirals detected here are within the cavity. 
Furthermore, there is no evidence for strong tilt of the inner (few au) disk of AB Aur
compared to the observed spirals, so this mechanism is unlikely to apply to our case.

Spiral patterns can 
also develop under external disturbances, like asymmetric accretion \citep{Hennebelle+etal_2016,Hennebelle+etal_2017}. 
Again, these studies assume a centrally condensed disk, different from the spirals detected here within the dust ring.
Indeed, for AB Aur, accretion from the outer envelope was invoked by 
\citet{tang2012} to explain the apparent counter-rotation of the gas 
in the outer (150--400 au) spirals. 
However, the opening angle of the induced spiral pattern is expected to 
be on the order of $h/r$ in the \citet{Hennebelle+etal_2016} model, again much smaller than our 0.3 radian value.

Thus, while our trailing spirals are highly suggestive of disturbances
due to planet-disk interaction or external disturbances, their pitch 
angle seems difficult to explain. We point out, however, that 
this apparent opening angle is only defined through a relatively
limited range of azimuth (less than 180$^\circ$). Furthermore, it 
may be affected by differential filtering of the extended emission 
at the various velocities (see section 3.2), especially near $v_{\rm sys}$ where the missing flux is largest.
The effect of missing flux on the morphology of the detected spirals can be studied by combining new ALMA observations with shorter baselines covering the size-scale of the spirals.

In summary, we report the detection of gaseous spirals within dust gap for the first time in the transitional disk AB Aur. In contrast to previous detections of the spirals in continuum emission, we are able to probe the kinematics of the spirals, which is mainly Keplerian. 
We argue that the spirals are most likely triggered by embedded object at two possible locations, either at 60-80 au and/or at around 30 au from the star. 

{\it Facilities:} \facility{ALMA}.

%
%
%

%
%
%

\acknowledgments{This paper makes use of the following ALMA data: ADS/JAO.ALMA\#2015.1.00889.S. ALMA is a partnership of ESO (representing its member states), NSF (USA) and NINS (Japan), together with NRC (Canada), NSC and ASIAA (Taiwan), and KASI (Republic of Korea), in cooperation with the Republic of Chile. The Joint ALMA Observatory is operated by ESO, AUI/NRAO and NAOJ. This research was partially supported by MOST grant MOST 105-2112-M-001-025-MY3. This research was supported in part by the "Programme National de Physique Stellaire" from INSU/CNRS, France.}
%
%
%
%
%
%
%
%
%

%
%
%
%
%

%
\bibliographystyle{apj}                       
\bibliography{abaur}

\end{document}